\documentclass[aps,prd,groupedaddress,nofootinbib,10pt,twocolumn]{revtex4-2}

\usepackage{dsfont}
\usepackage{natbib}
\usepackage{url}
\usepackage{color}
\usepackage{fullpage}
\usepackage[top=2cm, bottom=4.5cm, left=2.5cm, right=2.5cm]{geometry}
\usepackage{amsmath,amsthm,amsfonts,amssymb,amscd}
\usepackage{lastpage}
\usepackage{enumerate}
\usepackage[shortlabels]{enumitem}
\usepackage{fancyhdr}
\usepackage{mathrsfs}
\usepackage{xcolor}
\usepackage{listings}
\usepackage{dcolumn} 
\usepackage{bm}
\usepackage{graphicx}
\usepackage{hyperref}
\hypersetup{%
  colorlinks=true,
  linkcolor=blue,
  linkbordercolor={0 0 1}
}

\newcommand{\be}{\begin{equation}}
\newcommand{\ee}{\end{equation}}
\newcommand{\bea}{\begin{eqnarray}}
\newcommand{\eea}{\end{eqnarray}}
\newcommand{\nn}{\nonumber}

\newcommand{\lv}{\lambda_v}
\newcommand{\lp}{\lambda_+}
\newcommand{\lm}{\lambda_-}

\newcommand{\expect}[1]{\langle \widehat{#1} \rangle}
\newcommand{\rhm}{\dfrac{\rho}{\rho_{\text{max}}}}
\newcommand{\shm}{\dfrac{\sigma^2}{\sigma^2_{\text{max}}}}
\newcommand{\rbm}{\dfrac{\rho_b}{\rho_{\text{max}}}}
\newcommand{\sbm}{\dfrac{\sigma^2_b}{\sigma^2_{\text{max}}}}
\newcommand{\el}{e^{-\lv^2/s^2}}

\begin{document}

\title{An effective Friedmann equation for a bouncing anisotropic universe}

\author{Syed Moeez Hassan}
 \email{syed\_hassan@lums.edu.pk}
 \affiliation{%
 Department of Physics, Syed Babar Ali School of Science and Engineering, Lahore University of Management Sciences, Lahore 54792, Pakistan
}%

\date{\today}

\begin{abstract}

We derive generalized effective Friedmann and Raychaudhuri equations for a bouncing polymer quantized Bianchi-I spacetime. We further prove that the relation numerically derived in the literature between the matter energy density and the anisotropic shear at the bounce holds exactly for any type of matter content, and derive explicit expressions for the constants that appear in that relation.

\end{abstract}

\maketitle

\section{\label{sec-intro} Introduction}

Understanding the nature of singularities present in the early universe, and inside black holes is among one of the major challenges of fundamental theory in physics. The program of quantum gravity aims to address these challenges by quantizing the gravitational field. Among various proposals for such a theory, one major contender is loop quantum gravity (LQG) which provides a canonical approach towards the quantization of gravity \cite{Thiemann:2007pyv, Rovelli:2014ssa, Mehmood:2023jcn}. The central idea is that spacetime itself is discrete at a fundamental level. Loop quantum cosmology (LQC) is the application of these ideas to simpler cosmological spacetimes, with a key result being the resolution of the cosmic singularity into a bounce \cite{Ashtekar:2011ni, Agullo:2016tjh, Agullo_2023}. Both these theories, however, rely on heavy technical foundations, and use variables that at times obscure their classical connection.

As an alternate, polymer quantization is a scheme motivated from LQG and LQC, but distinct from them. It provides a simpler pathway to exploring the effects of a fundamental discreteness in spacetime by utilizing a Hilbert space with `discreteness built into it' \cite{Ashtekar:2002sn, Halvorson_2004}. This method has successfully been applied to cosmological models \cite{Husain:2003ry, BenAchour:2018jwq, Montani:2018uay, Barca:2019ane, Giovannetti:2019ewe, Giovannetti:2020nte, Giovannetti:2021bqh, Barca:2024qls}, gravitational collapse and black holes \cite{Ziprick:2016ogy, BenAchour:2018khr, Husain:2021ojz, Boldorini:2024bgg}, and even regular matter fields \cite{Corichi:2007tf, Husain:2007bj, Ashtekar:2002vh, Hossain:2009vd, Husain:2010gb, Laddha:2010hp, Hossain:2010eb, Hossain:2009ru, Hassan:2014sja, Hassan:2017cje, Ali:2017fhp}. In cosmology, although polymer quantization is distinct from LQC, several key results -- including a resolution of the initial big bang singularity into a bouncing universe -- are qualitatively similar \cite{Barca:2021qdn}.

While most work in cosmology has focused on the homogeneous and isotropic Friedmann-Lemaitre-Robertson-Walker (FLRW) model, the Bianchi spacetimes provide an anisotropic generalization of these models to a cosmological spacetime that is both simple enough to analyze, and with sufficient complexity to give rise to non-trivial behavior \cite{Schucker:2014wca, Campanelli:2006vb, Akarsu:2019pwn, Aluri:2022hzs, Hertzberg:2024uqy, Akarsu:2021max}. Furthermore, Bianchi spacetimes are also relevant to studying the initial `big-bang' singularity (and the anisotropic approach towards it) \cite{Misner:1969hg, Misner:1969ae, 1971SvPhU..13..745B}, and are becoming increasingly popular in light of the current Hubble tension \cite{Deliyergiyev:2025kun}.

In this work, we study the simplest Bianchi model: the Bianchi-I spacetime which features three different scale factors in three different coordinate directions. Both loop and polymer quantizations of this spacetime have been studied in detail \cite{Ashtekar:2009vc, deCesare:2019suk, Motaharfar:2023hil, Giovannetti:2021bqh}. Here, we polymer quantize the Bianchi-I spacetime, and obtain an effective Hamiltonian. From this Hamiltonian, we derive a generalized Friedmann equation for arbitrary matter content. We note that such an equation is not known for loop-quantized Bianchi-I models \cite{Motaharfar:2023hil, McNamara:2022dmf}, and to the best of our knowledge, this is the first such derivation in the literature. We further show that the relation between the matter energy density, and the anisotropic shear at the bounce first numerically proposed in \cite{McNamara:2022dmf} holds exactly for any type of matter content for the bouncing model considered here. Analytic expressions for the constants appearing in that relation are also derived. Throughout we work in natural units $c = \hbar = 8 \pi G = 1$.

\section{\label{sec-b1} Polymer quantized Bianchi-I spacetime}

We start with the Bianchi-I line element,
\be
ds^2=-N^2(t)dt^2+a_1^2(t)dx^2+a_2^2(t)dy^2+a_3^2(t)dz^2,
\ee
where $N(t)$ is the lapse function, and $a_i(t)$ are the three directional scale factors. In terms of these scale factors, the three directional Hubble parameters are defined as, 
\be
H_i=\frac{\dot{a_i}}{a_i},
\ee
with the mean Hubble parameter ($H$) given as,
\be
H = \frac{{1}}{3}\left(H_1+H_2+H_3\right) = \dfrac{\dot{v}}{3v},
\ee
where $v=a_1a_2a_3$ is the volume. The shear, which provides a measure of the anisotropy of spacetime is defined as,
\be
\sigma_i=H_i-H,
\ee
with the corresponding shear scalar given as,
\be
\sigma^2=\frac{1}{3}\left(\sigma_1^2+\sigma_2^2+\sigma_3^2\right).
\ee
Before polymerizing this spacetime, we transform to the volume and Misner anisotropy variables $(v,\beta_+, \beta_-)$ given as,
\bea \label{relation}
a_1(t) &=& v^{1/3} e^{\beta_+ +\sqrt{3}\beta_-},\nn \\
a_2(t) &=& v^{1/3} e^{\beta_+ -\sqrt{3}\beta_-},\nn \\
a_3(t) &=& v^{1/3} e^{-2\beta_+}.
\eea
In terms of these variables, and their canonically conjugate momenta $(p_v,p_+,p_-)$, the (ADM\footnote{Arnowitt--Deser--Misner (canonical Hamiltonian form of general relativity).}) Hamiltonian constraint takes the form\footnote{The diffeomorphism constraint vanishes identically due to homogeneity.},
\be
\mathcal{H} = H_v + H_+ + H_- + H_M \approx 0,
\ee
where
\be
H_v = -\dfrac{3}{16} vp_v^2, ~~~~ H_+ = \dfrac{p_+^2}{48v}, ~~~~ H_- = \dfrac{p_-^2}{48v},
\ee
and $H_M$ represents any (arbitrary) matter Hamiltonian.

We polymer quantize this Hamiltonian, following the method introduced in \cite{Zulfiqar:2025chv}, to obtain an effective Hamiltonian,
\bea
\label{eff-ham}
\mathcal{H_{\text{eff}}} &=& \expect{H_v} + \expect{H_+} + \expect{H_-} + \expect{H_M} \approx 0 \nn \\
&=& \frac{-3v}{32 \lambda_v^2} \left[1 - e^{-\lambda_v^2/s^2} \cos\left( 2{\lambda_v} p_v \right) \right] \nn \\
&+& \frac{v}{48\lambda_+^2} \sin^2\left(\frac{\lambda_{+}p_+}{v}\right) \nn \\
&+& \frac{v}{48\lambda_-^2} \sin^2\left(\frac{\lambda_{-}p_-}{v}\right) + H_M,
\eea
where, $\lambda_v,\lambda_\pm$ are the polymer scales for the volume and the anisotropies respectively, $s$ is the semiclassical state-width for the volume sector, and we have taken the semiclassical states for the anisotropy sector to be sharply peaked ($s_\pm \rightarrow \infty$). Polymer effects are contained in the sine and cosine terms, and we recover the classical theory in the limit that the discretization scales go to zero: $\lambda_v, \lambda_\pm \rightarrow 0$.

From this Hamiltonian constraint, we can write down an effective Friedmann-like equation,
\be
\label{eq-hub}
H^2 = \dfrac{\rho_{\text{tot}}}{12} \left(1 - \dfrac{\rho_{\text{tot}}}{\rho_0} \right) + f,
\ee
and an equation for the shear scalar as,
\be
\label{eq-sig}
\sigma^2 = \dfrac{\rho_+}{6} \left(1 - \dfrac{\rho_+}{\rho^+_0} \right) + \dfrac{\rho_-}{6} \left(1 - \dfrac{\rho_-}{\rho^-_0} \right),
\ee
where we have defined,\footnote{$f \rightarrow 0$ as the volume sector semi-classical state becomes sharply peaked ($s \rightarrow \infty$).}
\bea
\rho_{\text{tot}} = \rho_+ + \rho_- + \rho, ~~~~ f = \dfrac{e^{-2 \lv^2/s^2} - 1}{16^2 \lv^2}, \\
\rho = H_M/v, ~~~~ \rho_0 = 3/(16 \lv^2), \\
\rho_\pm = \expect{H_\pm}/v, ~~~~ \rho^\pm_0 = 1/(48 \lambda_\pm^2).
\eea

\section{\label{sec-frd} Generalized effective Friedmann equation}

To move towards writing down a generalized effective Friedmann equation for the polymerized Bianchi-I spacetime, we note that given the form of (\ref{eq-hub}, \ref{eq-sig}), it is not possible to cleanly separate out the anisotropy energy densities, and write down (\ref{eq-hub}) purely in terms of the matter energy density and the shear scalar. Indeed, a generalized effective Friedmann equation is not known for loop quantized Bianchi-I models. However, if we slightly simplify these equations, it turns out that writing down such an equation is possible, and furthermore, a unique relation exists between the shear scalar and the matter energy density at the bounce. In this section, we address the former, and return to the latter in the subsequent section.

We begin by defining the ratio of the two polymer anisotropy scales,
\be
l \equiv \dfrac{\lp}{\lm}.
\ee
Given this ratio, we assume that the two anisotropy momenta are related as,
\be
\label{passum}
p_- = l p_+.
\ee
This assumption implies that the two anisotropy energy densities are related as: $\rho_- = l^2 \rho_+$ (with the two being identical for same polymer scales: $l=1$). We note that since we are working with the Bianchi-I model, $p_\pm$ are constants of motion and hence can be freely specified in the initial data so as to satisfy (\ref{passum}) given some fixed value of $l$.

Using (\ref{passum}) in (\ref{eq-hub}, \ref{eq-sig}) then gives us,
\bea
H^2  &=& \dfrac{\rho}{12} \left(1-\dfrac{\rho}{\rho_0} \right) + \dfrac{S}{\rho_0} \sigma^2 + f \nn \\
&+& \dfrac{S}{12} \left( 1-2\dfrac{S+\rho}{\rho_0} \right) \left(1 \pm \sqrt{1 - \dfrac{12 \sigma^2}{S}} \right),
\eea
where we have defined,
\be
S = \dfrac{1}{96} \left( \dfrac{1}{\lp^2} + \dfrac{1}{\lm^2} \right).
\ee
Taking the isotropic limit ($\sigma^2 \rightarrow 0$) picks the negative sign in the square root above, and checking for the classical limit where $(\lv,\lp, \lm) \rightarrow 0$ fixes the ratios between the various polymer scales as,
\be
\label{ratio}
L^2 (1+l^2) = 9, ~~~ \text{with} ~~~ L \equiv \dfrac{\lv}{\lp}.
\ee

Putting all of the above together, we finally get the generalized effective Friedmann equation for the polymer quantized Bianchi-I spacetime,\footnote{The case when the anisotropy state widths ($s_\pm$) are also finite can be treated similarly, with the final result being,
\begin{math}
H^2 = \dfrac{1}{12} \rho \left(\sqrt{1 - \dfrac{24 (\sigma^2 - f_+ - f_-)}{\rho_0}} - \dfrac{\rho}{\rho_0} \right) \\
+ \dfrac{1}{2} (\sigma^2 - f_+ - f_-) + f,
\end{math}
with,\\ $f_\pm = (e^{-2 \lambda_\pm^2/v^2 s_\pm^2} - 1)/(48^2 \lambda_\pm^2)$. Note that this relation differs from (\ref{eff-fried}) in one significant aspect, which is that $f_\pm$ (unlike $f$) are explicitly dependent on the volume $v$, and are not constant.}
\footnote{In the classical limit ($\rho_0 \rightarrow \infty$), this expression reduces to,
\begin{math}
H^2 = \left( \dfrac{1}{12} \rho + \dfrac{1}{2} \sigma^2 \right) - \left( \dfrac{1}{12}\rho^2 + \rho \sigma^2 \right) \dfrac{1}{\rho_0} - \left( 6 \rho \sigma^4 \right) \dfrac{1}{\rho_0^2} \\
+ \mathcal{O}\left(1/\rho_0^3 \right),
\end{math}
and,\\ in the isotropic limit ($\sigma^2 \rightarrow 0$), we have,
\begin{math}
H^2 = \left( \dfrac{1}{12} \rho \left[ 1 - \dfrac{\rho}{\rho_0} \right] \right) + \left( \dfrac{1}{2}-\dfrac{\rho}{\rho_0} \right) \sigma^2 - \left( 6 \dfrac{\rho}{\rho_0^2} \right) \sigma^4 + \mathcal{O}\left( \sigma^6 \right),\\
\end{math}
where we have retained the first few leading order terms to show their explicit dependence.}
\be
\label{eff-fried}
H^2 = \dfrac{1}{12} \rho \left(\sqrt{1 - \dfrac{\sigma^2}{\sigma^2_{\text{max}}}} - \dfrac{\rho}{\rho_0} \right) + \dfrac{1}{2} \sigma^2 + f.
\ee
Using the same methods, we can also derive an effective Raychaudhuri equation,
\begin{widetext}
\be
\dot{H} = \dfrac{1}{N}\dfrac{dH}{dt} = -\dfrac{1}{8} \left( \sqrt{1 - \shm} - 2\dfrac{\rho}{\rho_0} \right) \left(\rho + P + 12 \sigma^2_{\text{max}} \sqrt{\dfrac{\sigma^2}{\sigma^2_{\text{max}}}} \sin^{-1} \left( \sqrt{\dfrac{\sigma^2}{\sigma^2_{\text{max}}}} \right) \right),
\ee
\end{widetext}
where $P \equiv -\partial H_M/\partial v$ is the standard matter pressure term, and $\sigma^2_{\text{max}}$ is a constant defined in (\ref{eq-smax}). It is tedious, but straightforward, to show that local matter energy conservation follows from the above two equations: $\dot{\rho} + 3H(\rho+P) = 0$.

\section{\label{sec-shear} Shear vs Matter energy density at the bounce}

We now turn to the computation of a relation between the anisotropic shear and the matter energy density at the bounce first numerically proposed in \cite{McNamara:2022dmf}. To start, we note from (\ref{eff-ham}), that the matter energy density is bounded above from,
\be
\rho_{\text{max}} = \dfrac{3}{32 \lambda_v^2} \left(1 + e^{-\lambda_v^2/s^2} \right) = \dfrac{\rho_0}{2} \left(1 + e^{-\lambda_v^2/s^2} \right)
\ee
(with $\rho_{\text{max}} = \rho_0$ when the volume state is sharply peaked), and from (\ref{eq-sig}), that the shear scalar is bounded above from,
\be
\label{eq-smax}
\sigma^2_{\text{max}} = \dfrac{1}{1152} \left( \dfrac{1}{\lp^2} + \dfrac{1}{\lm^2} \right) = \dfrac{\rho_0}{24}
\ee
(using (\ref{ratio}) in the last equality). Using these, we can re-write the effective Friedmann equation (\ref{eff-fried}) as,
\begin{widetext}
\be
(16 \lv H)^2 = 2(1+\el) \rhm \left[ \sqrt{1- \shm} - \dfrac{(1+\el)}{2} \rhm \right] + \shm + e^{-2\lambda_v^2/s^2} - 1.
\ee
\end{widetext}

To find a relation between the anisotropic shear scalar and the matter energy density at the bounce, we note that the bounce occurs at a turning point of the volume, where the mean Hubble parameter vanishes: $H=0$. Substituting this in the above and solving, we get,
\be
\label{eq-shear-rho}
\sbm = a \left( \rbm \right)^2 + b \left( \rbm \right) + c,
\ee
with,
\bea
a &=& - \left( 1+\el \right)^2,  \\
b &=& 2 (1+\el) (\el),  \\
c &=& 1 - e^{-2 \lv^2/s^2}.
\eea
In the limit that the semiclassical state is sharply peaked ($s \rightarrow \infty$), these constants become,
\be
\label{eq-const}
a = -4, ~~~ b = 4, ~~~ c = 0,
\ee
in agreement with earlier numerical results (Table-I in \cite{McNamara:2022dmf}).

\section{\label{sec-con} Discussion}

We have shown that with a reasonable assumption, effective Friedmann and Raychaudhuri equations can be derived for a bouncing anisotropic cosmological model. The model was chosen to be Bianchi-I, which is the simplest in the class of anisotropic cosmologies; and quantum gravitational effects were encoded through polymer quantization of this spacetime, where spacetime is assumed to be discrete at a fundamental level. This fundamental discreteness resolves the initial `big-bang' singularity, and gives rise to a bouncing universe. The anisotropic behavior is encoded in the shears, which remain bounded throughout.

The derivation presented here did not rely on any specific form of the matter Hamiltonian, apart from being minimally coupled to gravity (and that it has to be homogeneous to consistently couple with a homogeneous spacetime). In fact, we can even take the matter part to be quantized as well (and couple the expectation value of this matter Hamiltonian with gravity to obtain an effective matter Hamiltonian), with the effective Friedmann equation taking the same form (with matter energy density replaced with its effective counterpart). We further showed that from this effective Friedmann equation, a relation can be derived between the matter energy density, and the anisotropic shear at the bounce. The exact form of this relation, and values of the constants appearing in this relation were also derived, and it was shown that they produce the relation first numerically proposed in \cite{McNamara:2022dmf}.

A possible generalization of this approach is suggested by the form of (\ref{eq-hub}) and (\ref{eq-sig}). The goal is to eliminate the anisotropy energy densities appearing in (\ref{eq-hub}) by making use of (\ref{eq-sig}), and hence to re-write (\ref{eq-hub}) in terms of the matter energy density and the anisotropic shear alone. Classically, this is straightforward because the anisotropy energy densities appear linearly in both the shear and the Friedmann equation. However, in the bouncing model discussed here (just like LQC), we have non-linear combinations. Here, we were able to carry out this elimination by making the assumption (\ref{passum}), which is equivalent to assuming a linear relation between the anisotropy energy densities of the form $\rho_- = l^2 \rho_+$. It might be possible to relax this assumption by assuming a generic relation $\rho_- = f(\rho_+)$ (much like how the standard Friedmann equations in isotropic cosmology are supplemented by an equation of state relating the matter pressure and energy density), or even $f(\rho_-,\rho_+) = 0$. We leave such investigations for future work. Another direction is the extension to more complicated Bianchi models (like Bianchi-II or Bianchi-IX), which include a potential term for the anisotropies $(\beta_\pm$). In these models, given the presence of a potential, the anisotropy momenta are no longer constants of motion, and therefore, (\ref{passum}) will likely need to be generalized.

What do these results imply for other bouncing anisotropic models, and in particular for LQC? We note that our results strongly suggest the existence of an effective Friedmann equation, and in particular of a parabolic relation between the shear and matter energy density at the bounce, similar in form to (\ref{eff-fried}) and (\ref{eq-shear-rho}) respectively, for other bouncing models as well. For instance, the effective theory of deformation quantization bouncing models has been shown to be similar to polymer quantization \cite{Barca:2021epy}. For loop quantization, while it is known that LQC is not entirely equivalent to polymerized cosmologies \cite{Barca:2021qdn}, the match we obtain for the values of the constants (\ref{eq-const}) with those obtained in LQC \cite{McNamara:2022dmf} indicates the universality of the parabolic relation (\ref{eq-shear-rho}). Whether the small fluctuations away from these values seen in LQC are a numerical artifact, or a fundamental difference between LQC and polymerization warrants further study.

Overall, the results derived here present a significant step forward in the analysis of quantum-corrected (bouncing) anisotropic spacetimes, given that a generalized effective Friedmann equation is not available for loop (or similarly quantized) spacetimes. Typically, a Friedmann-like equation for such spacetimes is written in terms of the anisotropy energy densities (either separately, or as part of the `matter' energy density) \cite{Giovannetti:2021bqh, Barca:2025fpu, Barca:2021epy, Zulfiqar:2025chv}, and not in terms of the anisotropic shear. Where attempts have been made to write in terms of the shears (as noted in \cite{McNamara:2022dmf}), either the kinematic definition of the shear is changed \cite{Linsefors:2013bua}, or it is derived in the limit of small shear \cite{Chiou:2007sp}. The novel effective Friedmann equation presented here lends analytical insights into the nature of the bounce in Bianchi-I, and the interplay between the anisotropic shear and the matter energy density, especially in the deep-quantum regime.

\begin{acknowledgments}
This project was supported by the Higher Education Commission (HEC) of Pakistan through NRPU grant No. 20-15435.
\end{acknowledgments}

\bibliography{bib-bfes}

\begin{thebibliography}{51}%
\makeatletter
\providecommand \@ifxundefined [1]{%
 \@ifx{#1\undefined}
}%
\providecommand \@ifnum [1]{%
 \ifnum #1\expandafter \@firstoftwo
 \else \expandafter \@secondoftwo
 \fi
}%
\providecommand \@ifx [1]{%
 \ifx #1\expandafter \@firstoftwo
 \else \expandafter \@secondoftwo
 \fi
}%
\providecommand \natexlab [1]{#1}%
\providecommand \enquote  [1]{``#1''}%
\providecommand \bibnamefont  [1]{#1}%
\providecommand \bibfnamefont [1]{#1}%
\providecommand \citenamefont [1]{#1}%
\providecommand \href@noop [0]{\@secondoftwo}%
\providecommand \href [0]{\begingroup \@sanitize@url \@href}%
\providecommand \@href[1]{\@@startlink{#1}\@@href}%
\providecommand \@@href[1]{\endgroup#1\@@endlink}%
\providecommand \@sanitize@url [0]{\catcode `\\12\catcode `\$12\catcode `\&12\catcode `\#12\catcode `\^12\catcode `\_12\catcode `\%12\relax}%
\providecommand \@@startlink[1]{}%
\providecommand \@@endlink[0]{}%
\providecommand \url  [0]{\begingroup\@sanitize@url \@url }%
\providecommand \@url [1]{\endgroup\@href {#1}{\urlprefix }}%
\providecommand \urlprefix  [0]{URL }%
\providecommand \Eprint [0]{\href }%
\providecommand \doibase [0]{https://doi.org/}%
\providecommand \selectlanguage [0]{\@gobble}%
\providecommand \bibinfo  [0]{\@secondoftwo}%
\providecommand \bibfield  [0]{\@secondoftwo}%
\providecommand \translation [1]{[#1]}%
\providecommand \BibitemOpen [0]{}%
\providecommand \bibitemStop [0]{}%
\providecommand \bibitemNoStop [0]{.\EOS\space}%
\providecommand \EOS [0]{\spacefactor3000\relax}%
\providecommand \BibitemShut  [1]{\csname bibitem#1\endcsname}%
\let\auto@bib@innerbib\@empty
\bibitem [{\citenamefont {Thiemann}(2007)}]{Thiemann:2007pyv}%
  \BibitemOpen
  \bibfield  {author} {\bibinfo {author} {\bibfnamefont {T.}~\bibnamefont {Thiemann}},\ }\href {https://doi.org/10.1017/CBO9780511755682} {\emph {\bibinfo {title} {{Modern Canonical Quantum General Relativity}}}},\ Cambridge Monographs on Mathematical Physics\ (\bibinfo  {publisher} {Cambridge University Press},\ \bibinfo {year} {2007})\BibitemShut {NoStop}%
\bibitem [{\citenamefont {Rovelli}\ and\ \citenamefont {Vidotto}(2014)}]{Rovelli:2014ssa}%
  \BibitemOpen
  \bibfield  {author} {\bibinfo {author} {\bibfnamefont {C.}~\bibnamefont {Rovelli}}\ and\ \bibinfo {author} {\bibfnamefont {F.}~\bibnamefont {Vidotto}},\ }\href@noop {} {\emph {\bibinfo {title} {{Covariant Loop Quantum Gravity}: {An Elementary Introduction to Quantum Gravity and Spinfoam Theory}}}},\ Cambridge Monographs on Mathematical Physics\ (\bibinfo  {publisher} {Cambridge University Press},\ \bibinfo {year} {2014})\BibitemShut {NoStop}%
\bibitem [{\citenamefont {Mehmood}(2023)}]{Mehmood:2023jcn}%
  \BibitemOpen
  \bibfield  {author} {\bibinfo {author} {\bibfnamefont {H.}~\bibnamefont {Mehmood}},\ }\emph {\bibinfo {title} {{Quantum Gravity as a Theory of Connections}}},\ \href@noop {} {\bibinfo {type} {Other thesis}} (\bibinfo {year} {2023}),\ \Eprint {https://arxiv.org/abs/2309.16734} {arXiv:2309.16734 [gr-qc]} \BibitemShut {NoStop}%
\bibitem [{\citenamefont {Ashtekar}\ and\ \citenamefont {Singh}(2011)}]{Ashtekar:2011ni}%
  \BibitemOpen
  \bibfield  {author} {\bibinfo {author} {\bibfnamefont {A.}~\bibnamefont {Ashtekar}}\ and\ \bibinfo {author} {\bibfnamefont {P.}~\bibnamefont {Singh}},\ }\bibfield  {title} {\bibinfo {title} {{Loop Quantum Cosmology: A Status Report}},\ }\href {https://doi.org/10.1088/0264-9381/28/21/213001} {\bibfield  {journal} {\bibinfo  {journal} {Class. Quant. Grav.}\ }\textbf {\bibinfo {volume} {28}},\ \bibinfo {pages} {213001} (\bibinfo {year} {2011})},\ \Eprint {https://arxiv.org/abs/1108.0893} {arXiv:1108.0893 [gr-qc]} \BibitemShut {NoStop}%
\bibitem [{\citenamefont {Agullo}\ and\ \citenamefont {Singh}(2017)}]{Agullo:2016tjh}%
  \BibitemOpen
  \bibfield  {author} {\bibinfo {author} {\bibfnamefont {I.}~\bibnamefont {Agullo}}\ and\ \bibinfo {author} {\bibfnamefont {P.}~\bibnamefont {Singh}},\ }\bibinfo {title} {{Loop quantum cosmology.}},\ in\ \href {https://doi.org/10.1142/9789813220003_0007} {\emph {\bibinfo {booktitle} {{Loop Quantum Gravity}: {The First 30 Years}}}},\ \bibinfo {editor} {edited by\ \bibinfo {editor} {\bibfnamefont {A.}~\bibnamefont {Ashtekar}}\ and\ \bibinfo {editor} {\bibfnamefont {J.}~\bibnamefont {Pullin}}}\ (\bibinfo  {publisher} {WSP},\ \bibinfo {year} {2017})\ pp.\ \bibinfo {pages} {183--240},\ \Eprint {https://arxiv.org/abs/1612.01236} {arXiv:1612.01236 [gr-qc]} \BibitemShut {NoStop}%
\bibitem [{\citenamefont {Agullo}\ \emph {et~al.}(2023)\citenamefont {Agullo}, \citenamefont {Wang},\ and\ \citenamefont {Wilson-Ewing}}]{Agullo_2023}%
  \BibitemOpen
  \bibfield  {author} {\bibinfo {author} {\bibfnamefont {I.}~\bibnamefont {Agullo}}, \bibinfo {author} {\bibfnamefont {A.}~\bibnamefont {Wang}},\ and\ \bibinfo {author} {\bibfnamefont {E.}~\bibnamefont {Wilson-Ewing}},\ }\bibinfo {title} {Loop quantum cosmology: Relation between theory and observations},\ in\ \href {https://doi.org/10.1007/978-981-19-3079-9_103-1} {\emph {\bibinfo {booktitle} {Handbook of Quantum Gravity}}}\ (\bibinfo  {publisher} {Springer Nature Singapore},\ \bibinfo {year} {2023})\ p.\ \bibinfo {pages} {1–46}\BibitemShut {NoStop}%
\bibitem [{\citenamefont {Ashtekar}\ \emph {et~al.}(2003{\natexlab{a}})\citenamefont {Ashtekar}, \citenamefont {Fairhurst},\ and\ \citenamefont {Willis}}]{Ashtekar:2002sn}%
  \BibitemOpen
  \bibfield  {author} {\bibinfo {author} {\bibfnamefont {A.}~\bibnamefont {Ashtekar}}, \bibinfo {author} {\bibfnamefont {S.}~\bibnamefont {Fairhurst}},\ and\ \bibinfo {author} {\bibfnamefont {J.~L.}\ \bibnamefont {Willis}},\ }\bibfield  {title} {\bibinfo {title} {{Quantum gravity, shadow states, and quantum mechanics}},\ }\href {https://doi.org/10.1088/0264-9381/20/6/302} {\bibfield  {journal} {\bibinfo  {journal} {Class. Quant. Grav.}\ }\textbf {\bibinfo {volume} {20}},\ \bibinfo {pages} {1031} (\bibinfo {year} {2003}{\natexlab{a}})},\ \Eprint {https://arxiv.org/abs/gr-qc/0207106} {arXiv:gr-qc/0207106} \BibitemShut {NoStop}%
\bibitem [{\citenamefont {Halvorson}(2004)}]{Halvorson_2004}%
  \BibitemOpen
  \bibfield  {author} {\bibinfo {author} {\bibfnamefont {H.}~\bibnamefont {Halvorson}},\ }\bibfield  {title} {\bibinfo {title} {Complementarity of representations in quantum mechanics},\ }\href {https://doi.org/10.1016/j.shpsb.2003.01.001} {\bibfield  {journal} {\bibinfo  {journal} {Studies in History and Philosophy of Science Part B: Studies in History and Philosophy of Modern Physics}\ }\textbf {\bibinfo {volume} {35}},\ \bibinfo {pages} {45–56} (\bibinfo {year} {2004})}\BibitemShut {NoStop}%
\bibitem [{\citenamefont {Husain}\ and\ \citenamefont {Winkler}(2004)}]{Husain:2003ry}%
  \BibitemOpen
  \bibfield  {author} {\bibinfo {author} {\bibfnamefont {V.}~\bibnamefont {Husain}}\ and\ \bibinfo {author} {\bibfnamefont {O.}~\bibnamefont {Winkler}},\ }\bibfield  {title} {\bibinfo {title} {{On singularity resolution in quantum gravity}},\ }\href {https://doi.org/10.1103/PhysRevD.69.084016} {\bibfield  {journal} {\bibinfo  {journal} {Phys. Rev. D}\ }\textbf {\bibinfo {volume} {69}},\ \bibinfo {pages} {084016} (\bibinfo {year} {2004})},\ \Eprint {https://arxiv.org/abs/gr-qc/0312094} {arXiv:gr-qc/0312094} \BibitemShut {NoStop}%
\bibitem [{\citenamefont {Ben~Achour}\ and\ \citenamefont {Livine}(2019)}]{BenAchour:2018jwq}%
  \BibitemOpen
  \bibfield  {author} {\bibinfo {author} {\bibfnamefont {J.}~\bibnamefont {Ben~Achour}}\ and\ \bibinfo {author} {\bibfnamefont {E.~R.}\ \bibnamefont {Livine}},\ }\bibfield  {title} {\bibinfo {title} {{Polymer Quantum Cosmology: Lifting quantization ambiguities using a $SL(2,\mathbb{R})$ conformal symmetry}},\ }\href {https://doi.org/10.1103/PhysRevD.99.126013} {\bibfield  {journal} {\bibinfo  {journal} {Phys. Rev. D}\ }\textbf {\bibinfo {volume} {99}},\ \bibinfo {pages} {126013} (\bibinfo {year} {2019})},\ \Eprint {https://arxiv.org/abs/1806.09290} {arXiv:1806.09290 [gr-qc]} \BibitemShut {NoStop}%
\bibitem [{\citenamefont {Montani}\ \emph {et~al.}(2019)\citenamefont {Montani}, \citenamefont {Mantero}, \citenamefont {Bombacigno}, \citenamefont {Cianfrani},\ and\ \citenamefont {Barca}}]{Montani:2018uay}%
  \BibitemOpen
  \bibfield  {author} {\bibinfo {author} {\bibfnamefont {G.}~\bibnamefont {Montani}}, \bibinfo {author} {\bibfnamefont {C.}~\bibnamefont {Mantero}}, \bibinfo {author} {\bibfnamefont {F.}~\bibnamefont {Bombacigno}}, \bibinfo {author} {\bibfnamefont {F.}~\bibnamefont {Cianfrani}},\ and\ \bibinfo {author} {\bibfnamefont {G.}~\bibnamefont {Barca}},\ }\bibfield  {title} {\bibinfo {title} {{Semiclassical and quantum analysis of the isotropic Universe in the polymer paradigm}},\ }\href {https://doi.org/10.1103/PhysRevD.99.063534} {\bibfield  {journal} {\bibinfo  {journal} {Phys. Rev. D}\ }\textbf {\bibinfo {volume} {99}},\ \bibinfo {pages} {063534} (\bibinfo {year} {2019})},\ \Eprint {https://arxiv.org/abs/1806.10364} {arXiv:1806.10364 [gr-qc]} \BibitemShut {NoStop}%
\bibitem [{\citenamefont {Barca}\ \emph {et~al.}(2019)\citenamefont {Barca}, \citenamefont {Di~Antonio}, \citenamefont {Montani},\ and\ \citenamefont {Patti}}]{Barca:2019ane}%
  \BibitemOpen
  \bibfield  {author} {\bibinfo {author} {\bibfnamefont {G.}~\bibnamefont {Barca}}, \bibinfo {author} {\bibfnamefont {P.}~\bibnamefont {Di~Antonio}}, \bibinfo {author} {\bibfnamefont {G.}~\bibnamefont {Montani}},\ and\ \bibinfo {author} {\bibfnamefont {A.}~\bibnamefont {Patti}},\ }\bibfield  {title} {\bibinfo {title} {{Semiclassical and quantum polymer effects in a flat isotropic universe}},\ }\href {https://doi.org/10.1103/PhysRevD.99.123509} {\bibfield  {journal} {\bibinfo  {journal} {Phys. Rev. D}\ }\textbf {\bibinfo {volume} {99}},\ \bibinfo {pages} {123509} (\bibinfo {year} {2019})},\ \Eprint {https://arxiv.org/abs/1902.02128} {arXiv:1902.02128 [gr-qc]} \BibitemShut {NoStop}%
\bibitem [{\citenamefont {Giovannetti}\ and\ \citenamefont {Montani}(2019)}]{Giovannetti:2019ewe}%
  \BibitemOpen
  \bibfield  {author} {\bibinfo {author} {\bibfnamefont {E.}~\bibnamefont {Giovannetti}}\ and\ \bibinfo {author} {\bibfnamefont {G.}~\bibnamefont {Montani}},\ }\bibfield  {title} {\bibinfo {title} {{Polymer representation of the Bianchi IX Cosmology in the Misner variables}},\ }\href {https://doi.org/10.1103/PhysRevD.100.104058} {\bibfield  {journal} {\bibinfo  {journal} {Phys. Rev. D}\ }\textbf {\bibinfo {volume} {100}},\ \bibinfo {pages} {104058} (\bibinfo {year} {2019})},\ \Eprint {https://arxiv.org/abs/1907.12083} {arXiv:1907.12083 [gr-qc]} \BibitemShut {NoStop}%
\bibitem [{\citenamefont {Giovannetti}\ \emph {et~al.}(2022{\natexlab{a}})\citenamefont {Giovannetti}, \citenamefont {Barca}, \citenamefont {Mandini},\ and\ \citenamefont {Montani}}]{Giovannetti:2020nte}%
  \BibitemOpen
  \bibfield  {author} {\bibinfo {author} {\bibfnamefont {E.}~\bibnamefont {Giovannetti}}, \bibinfo {author} {\bibfnamefont {G.}~\bibnamefont {Barca}}, \bibinfo {author} {\bibfnamefont {F.}~\bibnamefont {Mandini}},\ and\ \bibinfo {author} {\bibfnamefont {G.}~\bibnamefont {Montani}},\ }\bibfield  {title} {\bibinfo {title} {{Polymer Dynamics of Isotropic Universe in Ashtekar and in Volume Variables}},\ }\href {https://doi.org/10.3390/universe8060302} {\bibfield  {journal} {\bibinfo  {journal} {Universe}\ }\textbf {\bibinfo {volume} {8}},\ \bibinfo {pages} {302} (\bibinfo {year} {2022}{\natexlab{a}})},\ \Eprint {https://arxiv.org/abs/2006.10614} {arXiv:2006.10614 [gr-qc]} \BibitemShut {NoStop}%
\bibitem [{\citenamefont {Giovannetti}\ \emph {et~al.}(2022{\natexlab{b}})\citenamefont {Giovannetti}, \citenamefont {Montani},\ and\ \citenamefont {Schiattarella}}]{Giovannetti:2021bqh}%
  \BibitemOpen
  \bibfield  {author} {\bibinfo {author} {\bibfnamefont {E.}~\bibnamefont {Giovannetti}}, \bibinfo {author} {\bibfnamefont {G.}~\bibnamefont {Montani}},\ and\ \bibinfo {author} {\bibfnamefont {S.}~\bibnamefont {Schiattarella}},\ }\bibfield  {title} {\bibinfo {title} {{Semiclassical and quantum features of the Bianchi I cosmology in the polymer representation}},\ }\href {https://doi.org/10.1103/PhysRevD.105.064011} {\bibfield  {journal} {\bibinfo  {journal} {Phys. Rev. D}\ }\textbf {\bibinfo {volume} {105}},\ \bibinfo {pages} {064011} (\bibinfo {year} {2022}{\natexlab{b}})},\ \Eprint {https://arxiv.org/abs/2105.00360} {arXiv:2105.00360 [gr-qc]} \BibitemShut {NoStop}%
\bibitem [{\citenamefont {Barca}\ \emph {et~al.}(2024)\citenamefont {Barca}, \citenamefont {Boglioni},\ and\ \citenamefont {Montani}}]{Barca:2024qls}%
  \BibitemOpen
  \bibfield  {author} {\bibinfo {author} {\bibfnamefont {G.}~\bibnamefont {Barca}}, \bibinfo {author} {\bibfnamefont {L.}~\bibnamefont {Boglioni}},\ and\ \bibinfo {author} {\bibfnamefont {G.}~\bibnamefont {Montani}},\ }\bibfield  {title} {\bibinfo {title} {{Quantum isotropic Universe in RQM analogy: The cosmological horizon}},\ }\href {https://doi.org/10.1016/j.dark.2024.101540} {\bibfield  {journal} {\bibinfo  {journal} {Phys. Dark Univ.}\ }\textbf {\bibinfo {volume} {45}},\ \bibinfo {pages} {101540} (\bibinfo {year} {2024})},\ \Eprint {https://arxiv.org/abs/2404.07056} {arXiv:2404.07056 [gr-qc]} \BibitemShut {NoStop}%
\bibitem [{\citenamefont {Ziprick}\ \emph {et~al.}(2016)\citenamefont {Ziprick}, \citenamefont {Gegenberg},\ and\ \citenamefont {Kunstatter}}]{Ziprick:2016ogy}%
  \BibitemOpen
  \bibfield  {author} {\bibinfo {author} {\bibfnamefont {J.}~\bibnamefont {Ziprick}}, \bibinfo {author} {\bibfnamefont {J.}~\bibnamefont {Gegenberg}},\ and\ \bibinfo {author} {\bibfnamefont {G.}~\bibnamefont {Kunstatter}},\ }\bibfield  {title} {\bibinfo {title} {{Polymer Quantization of a Self-Gravitating Thin Shell}},\ }\href {https://doi.org/10.1103/PhysRevD.94.104076} {\bibfield  {journal} {\bibinfo  {journal} {Phys. Rev. D}\ }\textbf {\bibinfo {volume} {94}},\ \bibinfo {pages} {104076} (\bibinfo {year} {2016})},\ \Eprint {https://arxiv.org/abs/1609.06665} {arXiv:1609.06665 [gr-qc]} \BibitemShut {NoStop}%
\bibitem [{\citenamefont {Ben~Achour}\ \emph {et~al.}(2018)\citenamefont {Ben~Achour}, \citenamefont {Lamy}, \citenamefont {Liu},\ and\ \citenamefont {Noui}}]{BenAchour:2018khr}%
  \BibitemOpen
  \bibfield  {author} {\bibinfo {author} {\bibfnamefont {J.}~\bibnamefont {Ben~Achour}}, \bibinfo {author} {\bibfnamefont {F.}~\bibnamefont {Lamy}}, \bibinfo {author} {\bibfnamefont {H.}~\bibnamefont {Liu}},\ and\ \bibinfo {author} {\bibfnamefont {K.}~\bibnamefont {Noui}},\ }\bibfield  {title} {\bibinfo {title} {{Polymer Schwarzschild black hole: An effective metric}},\ }\href {https://doi.org/10.1209/0295-5075/123/20006} {\bibfield  {journal} {\bibinfo  {journal} {EPL}\ }\textbf {\bibinfo {volume} {123}},\ \bibinfo {pages} {20006} (\bibinfo {year} {2018})},\ \Eprint {https://arxiv.org/abs/1803.01152} {arXiv:1803.01152 [gr-qc]} \BibitemShut {NoStop}%
\bibitem [{\citenamefont {Husain}\ \emph {et~al.}(2022)\citenamefont {Husain}, \citenamefont {Kelly}, \citenamefont {Santacruz},\ and\ \citenamefont {Wilson-Ewing}}]{Husain:2021ojz}%
  \BibitemOpen
  \bibfield  {author} {\bibinfo {author} {\bibfnamefont {V.}~\bibnamefont {Husain}}, \bibinfo {author} {\bibfnamefont {J.~G.}\ \bibnamefont {Kelly}}, \bibinfo {author} {\bibfnamefont {R.}~\bibnamefont {Santacruz}},\ and\ \bibinfo {author} {\bibfnamefont {E.}~\bibnamefont {Wilson-Ewing}},\ }\bibfield  {title} {\bibinfo {title} {{Quantum Gravity of Dust Collapse: Shock Waves from Black Holes}},\ }\href {https://doi.org/10.1103/PhysRevLett.128.121301} {\bibfield  {journal} {\bibinfo  {journal} {Phys. Rev. Lett.}\ }\textbf {\bibinfo {volume} {128}},\ \bibinfo {pages} {121301} (\bibinfo {year} {2022})},\ \Eprint {https://arxiv.org/abs/2109.08667} {arXiv:2109.08667 [gr-qc]} \BibitemShut {NoStop}%
\bibitem [{\citenamefont {Boldorini}\ and\ \citenamefont {Montani}(2024)}]{Boldorini:2024bgg}%
  \BibitemOpen
  \bibfield  {author} {\bibinfo {author} {\bibfnamefont {L.}~\bibnamefont {Boldorini}}\ and\ \bibinfo {author} {\bibfnamefont {G.}~\bibnamefont {Montani}},\ }\bibfield  {title} {\bibinfo {title} {{Effective quantum gravitational collapse in a polymer framework}},\ }\href {https://doi.org/10.1088/1475-7516/2024/10/090} {\bibfield  {journal} {\bibinfo  {journal} {JCAP}\ }\textbf {\bibinfo {volume} {10}},\ \bibinfo {pages} {090}},\ \Eprint {https://arxiv.org/abs/2406.03279} {arXiv:2406.03279 [gr-qc]} \BibitemShut {NoStop}%
\bibitem [{\citenamefont {Corichi}\ \emph {et~al.}(2007)\citenamefont {Corichi}, \citenamefont {Vukasinac},\ and\ \citenamefont {Zapata}}]{Corichi:2007tf}%
  \BibitemOpen
  \bibfield  {author} {\bibinfo {author} {\bibfnamefont {A.}~\bibnamefont {Corichi}}, \bibinfo {author} {\bibfnamefont {T.}~\bibnamefont {Vukasinac}},\ and\ \bibinfo {author} {\bibfnamefont {J.~A.}\ \bibnamefont {Zapata}},\ }\bibfield  {title} {\bibinfo {title} {{Polymer Quantum Mechanics and its Continuum Limit}},\ }\href {https://doi.org/10.1103/PhysRevD.76.044016} {\bibfield  {journal} {\bibinfo  {journal} {Phys. Rev. D}\ }\textbf {\bibinfo {volume} {76}},\ \bibinfo {pages} {044016} (\bibinfo {year} {2007})},\ \Eprint {https://arxiv.org/abs/0704.0007} {arXiv:0704.0007 [gr-qc]} \BibitemShut {NoStop}%
\bibitem [{\citenamefont {Husain}\ \emph {et~al.}(2007)\citenamefont {Husain}, \citenamefont {Louko},\ and\ \citenamefont {Winkler}}]{Husain:2007bj}%
  \BibitemOpen
  \bibfield  {author} {\bibinfo {author} {\bibfnamefont {V.}~\bibnamefont {Husain}}, \bibinfo {author} {\bibfnamefont {J.}~\bibnamefont {Louko}},\ and\ \bibinfo {author} {\bibfnamefont {O.}~\bibnamefont {Winkler}},\ }\bibfield  {title} {\bibinfo {title} {{Quantum gravity and the Coulomb potential}},\ }\href {https://doi.org/10.1103/PhysRevD.76.084002} {\bibfield  {journal} {\bibinfo  {journal} {Phys. Rev. D}\ }\textbf {\bibinfo {volume} {76}},\ \bibinfo {pages} {084002} (\bibinfo {year} {2007})},\ \Eprint {https://arxiv.org/abs/0707.0273} {arXiv:0707.0273 [gr-qc]} \BibitemShut {NoStop}%
\bibitem [{\citenamefont {Ashtekar}\ \emph {et~al.}(2003{\natexlab{b}})\citenamefont {Ashtekar}, \citenamefont {Lewandowski},\ and\ \citenamefont {Sahlmann}}]{Ashtekar:2002vh}%
  \BibitemOpen
  \bibfield  {author} {\bibinfo {author} {\bibfnamefont {A.}~\bibnamefont {Ashtekar}}, \bibinfo {author} {\bibfnamefont {J.}~\bibnamefont {Lewandowski}},\ and\ \bibinfo {author} {\bibfnamefont {H.}~\bibnamefont {Sahlmann}},\ }\bibfield  {title} {\bibinfo {title} {{Polymer and Fock representations for a scalar field}},\ }\href {https://doi.org/10.1088/0264-9381/20/1/103} {\bibfield  {journal} {\bibinfo  {journal} {Class. Quant. Grav.}\ }\textbf {\bibinfo {volume} {20}},\ \bibinfo {pages} {L11} (\bibinfo {year} {2003}{\natexlab{b}})},\ \Eprint {https://arxiv.org/abs/gr-qc/0211012} {arXiv:gr-qc/0211012} \BibitemShut {NoStop}%
\bibitem [{\citenamefont {Hossain}\ \emph {et~al.}(2009)\citenamefont {Hossain}, \citenamefont {Husain},\ and\ \citenamefont {Seahra}}]{Hossain:2009vd}%
  \BibitemOpen
  \bibfield  {author} {\bibinfo {author} {\bibfnamefont {G.~M.}\ \bibnamefont {Hossain}}, \bibinfo {author} {\bibfnamefont {V.}~\bibnamefont {Husain}},\ and\ \bibinfo {author} {\bibfnamefont {S.~S.}\ \bibnamefont {Seahra}},\ }\bibfield  {title} {\bibinfo {title} {{Background independent quantization and wave propagation}},\ }\href {https://doi.org/10.1103/PhysRevD.80.044018} {\bibfield  {journal} {\bibinfo  {journal} {Phys. Rev. D}\ }\textbf {\bibinfo {volume} {80}},\ \bibinfo {pages} {044018} (\bibinfo {year} {2009})},\ \Eprint {https://arxiv.org/abs/0906.4046} {arXiv:0906.4046 [hep-th]} \BibitemShut {NoStop}%
\bibitem [{\citenamefont {Husain}\ and\ \citenamefont {Kreienbuehl}(2010)}]{Husain:2010gb}%
  \BibitemOpen
  \bibfield  {author} {\bibinfo {author} {\bibfnamefont {V.}~\bibnamefont {Husain}}\ and\ \bibinfo {author} {\bibfnamefont {A.}~\bibnamefont {Kreienbuehl}},\ }\bibfield  {title} {\bibinfo {title} {{Ultraviolet behavior in background independent quantum field theory}},\ }\href {https://doi.org/10.1103/PhysRevD.81.084043} {\bibfield  {journal} {\bibinfo  {journal} {Phys. Rev. D}\ }\textbf {\bibinfo {volume} {81}},\ \bibinfo {pages} {084043} (\bibinfo {year} {2010})},\ \Eprint {https://arxiv.org/abs/1002.0138} {arXiv:1002.0138 [gr-qc]} \BibitemShut {NoStop}%
\bibitem [{\citenamefont {Laddha}\ and\ \citenamefont {Varadarajan}(2010)}]{Laddha:2010hp}%
  \BibitemOpen
  \bibfield  {author} {\bibinfo {author} {\bibfnamefont {A.}~\bibnamefont {Laddha}}\ and\ \bibinfo {author} {\bibfnamefont {M.}~\bibnamefont {Varadarajan}},\ }\bibfield  {title} {\bibinfo {title} {{Polymer quantization of the free scalar field and its classical limit}},\ }\href {https://doi.org/10.1088/0264-9381/27/17/175010} {\bibfield  {journal} {\bibinfo  {journal} {Class. Quant. Grav.}\ }\textbf {\bibinfo {volume} {27}},\ \bibinfo {pages} {175010} (\bibinfo {year} {2010})},\ \Eprint {https://arxiv.org/abs/1001.3505} {arXiv:1001.3505 [gr-qc]} \BibitemShut {NoStop}%
\bibitem [{\citenamefont {Hossain}\ \emph {et~al.}(2010{\natexlab{a}})\citenamefont {Hossain}, \citenamefont {Husain},\ and\ \citenamefont {Seahra}}]{Hossain:2010eb}%
  \BibitemOpen
  \bibfield  {author} {\bibinfo {author} {\bibfnamefont {G.~M.}\ \bibnamefont {Hossain}}, \bibinfo {author} {\bibfnamefont {V.}~\bibnamefont {Husain}},\ and\ \bibinfo {author} {\bibfnamefont {S.~S.}\ \bibnamefont {Seahra}},\ }\bibfield  {title} {\bibinfo {title} {{The Propagator in polymer quantum field theory}},\ }\href {https://doi.org/10.1103/PhysRevD.82.124032} {\bibfield  {journal} {\bibinfo  {journal} {Phys. Rev. D}\ }\textbf {\bibinfo {volume} {82}},\ \bibinfo {pages} {124032} (\bibinfo {year} {2010}{\natexlab{a}})},\ \Eprint {https://arxiv.org/abs/1007.5500} {arXiv:1007.5500 [gr-qc]} \BibitemShut {NoStop}%
\bibitem [{\citenamefont {Hossain}\ \emph {et~al.}(2010{\natexlab{b}})\citenamefont {Hossain}, \citenamefont {Husain},\ and\ \citenamefont {Seahra}}]{Hossain:2009ru}%
  \BibitemOpen
  \bibfield  {author} {\bibinfo {author} {\bibfnamefont {G.~M.}\ \bibnamefont {Hossain}}, \bibinfo {author} {\bibfnamefont {V.}~\bibnamefont {Husain}},\ and\ \bibinfo {author} {\bibfnamefont {S.~S.}\ \bibnamefont {Seahra}},\ }\bibfield  {title} {\bibinfo {title} {{Non-singular inflationary universe from polymer matter}},\ }\href {https://doi.org/10.1103/PhysRevD.81.024005} {\bibfield  {journal} {\bibinfo  {journal} {Phys. Rev. D}\ }\textbf {\bibinfo {volume} {81}},\ \bibinfo {pages} {024005} (\bibinfo {year} {2010}{\natexlab{b}})},\ \Eprint {https://arxiv.org/abs/0906.2798} {arXiv:0906.2798 [astro-ph.CO]} \BibitemShut {NoStop}%
\bibitem [{\citenamefont {Hassan}\ \emph {et~al.}(2015)\citenamefont {Hassan}, \citenamefont {Husain},\ and\ \citenamefont {Seahra}}]{Hassan:2014sja}%
  \BibitemOpen
  \bibfield  {author} {\bibinfo {author} {\bibfnamefont {S.~M.}\ \bibnamefont {Hassan}}, \bibinfo {author} {\bibfnamefont {V.}~\bibnamefont {Husain}},\ and\ \bibinfo {author} {\bibfnamefont {S.~S.}\ \bibnamefont {Seahra}},\ }\bibfield  {title} {\bibinfo {title} {{Polymer inflation}},\ }\href {https://doi.org/10.1103/PhysRevD.91.065006} {\bibfield  {journal} {\bibinfo  {journal} {Phys. Rev. D}\ }\textbf {\bibinfo {volume} {91}},\ \bibinfo {pages} {065006} (\bibinfo {year} {2015})},\ \Eprint {https://arxiv.org/abs/1409.6218} {arXiv:1409.6218 [astro-ph.CO]} \BibitemShut {NoStop}%
\bibitem [{\citenamefont {Hassan}\ and\ \citenamefont {Husain}(2017)}]{Hassan:2017cje}%
  \BibitemOpen
  \bibfield  {author} {\bibinfo {author} {\bibfnamefont {S.~M.}\ \bibnamefont {Hassan}}\ and\ \bibinfo {author} {\bibfnamefont {V.}~\bibnamefont {Husain}},\ }\bibfield  {title} {\bibinfo {title} {{Semiclassical cosmology with polymer matter}},\ }\href {https://doi.org/10.1088/1361-6382/aa6455} {\bibfield  {journal} {\bibinfo  {journal} {Class. Quant. Grav.}\ }\textbf {\bibinfo {volume} {34}},\ \bibinfo {pages} {084003} (\bibinfo {year} {2017})},\ \Eprint {https://arxiv.org/abs/1705.00398} {arXiv:1705.00398 [gr-qc]} \BibitemShut {NoStop}%
\bibitem [{\citenamefont {Ali}\ and\ \citenamefont {Seahra}(2017)}]{Ali:2017fhp}%
  \BibitemOpen
  \bibfield  {author} {\bibinfo {author} {\bibfnamefont {M.}~\bibnamefont {Ali}}\ and\ \bibinfo {author} {\bibfnamefont {S.~S.}\ \bibnamefont {Seahra}},\ }\bibfield  {title} {\bibinfo {title} {{Natural Inflation from Polymer Quantization}},\ }\href {https://doi.org/10.1103/PhysRevD.96.103524} {\bibfield  {journal} {\bibinfo  {journal} {Phys. Rev. D}\ }\textbf {\bibinfo {volume} {96}},\ \bibinfo {pages} {103524} (\bibinfo {year} {2017})},\ \Eprint {https://arxiv.org/abs/1709.03960} {arXiv:1709.03960 [gr-qc]} \BibitemShut {NoStop}%
\bibitem [{\citenamefont {Barca}\ \emph {et~al.}(2021)\citenamefont {Barca}, \citenamefont {Giovannetti},\ and\ \citenamefont {Montani}}]{Barca:2021qdn}%
  \BibitemOpen
  \bibfield  {author} {\bibinfo {author} {\bibfnamefont {G.}~\bibnamefont {Barca}}, \bibinfo {author} {\bibfnamefont {E.}~\bibnamefont {Giovannetti}},\ and\ \bibinfo {author} {\bibfnamefont {G.}~\bibnamefont {Montani}},\ }\bibfield  {title} {\bibinfo {title} {{An Overview on the Nature of the Bounce in LQC and PQM}},\ }\href {https://doi.org/10.3390/universe7090327} {\bibfield  {journal} {\bibinfo  {journal} {Universe}\ }\textbf {\bibinfo {volume} {7}},\ \bibinfo {pages} {327} (\bibinfo {year} {2021})},\ \Eprint {https://arxiv.org/abs/2109.08645} {arXiv:2109.08645 [gr-qc]} \BibitemShut {NoStop}%
\bibitem [{\citenamefont {Schucker}\ \emph {et~al.}(2014)\citenamefont {Schucker}, \citenamefont {Tilquin},\ and\ \citenamefont {Valent}}]{Schucker:2014wca}%
  \BibitemOpen
  \bibfield  {author} {\bibinfo {author} {\bibfnamefont {T.}~\bibnamefont {Schucker}}, \bibinfo {author} {\bibfnamefont {A.}~\bibnamefont {Tilquin}},\ and\ \bibinfo {author} {\bibfnamefont {G.}~\bibnamefont {Valent}},\ }\bibfield  {title} {\bibinfo {title} {{Bianchi I meets the Hubble diagram}},\ }\href {https://doi.org/10.1093/mnras/stu1656} {\bibfield  {journal} {\bibinfo  {journal} {Mon. Not. Roy. Astron. Soc.}\ }\textbf {\bibinfo {volume} {444}},\ \bibinfo {pages} {2820} (\bibinfo {year} {2014})},\ \Eprint {https://arxiv.org/abs/1405.6523} {arXiv:1405.6523 [astro-ph.CO]} \BibitemShut {NoStop}%
\bibitem [{\citenamefont {Campanelli}\ \emph {et~al.}(2006)\citenamefont {Campanelli}, \citenamefont {Cea},\ and\ \citenamefont {Tedesco}}]{Campanelli:2006vb}%
  \BibitemOpen
  \bibfield  {author} {\bibinfo {author} {\bibfnamefont {L.}~\bibnamefont {Campanelli}}, \bibinfo {author} {\bibfnamefont {P.}~\bibnamefont {Cea}},\ and\ \bibinfo {author} {\bibfnamefont {L.}~\bibnamefont {Tedesco}},\ }\bibfield  {title} {\bibinfo {title} {{Ellipsoidal Universe Can Solve The CMB Quadrupole Problem}},\ }\href {https://doi.org/10.1103/PhysRevLett.97.131302} {\bibfield  {journal} {\bibinfo  {journal} {Phys. Rev. Lett.}\ }\textbf {\bibinfo {volume} {97}},\ \bibinfo {pages} {131302} (\bibinfo {year} {2006})},\ \bibinfo {note} {[Erratum: Phys.Rev.Lett. 97, 209903 (2006)]},\ \Eprint {https://arxiv.org/abs/astro-ph/0606266} {arXiv:astro-ph/0606266} \BibitemShut {NoStop}%
\bibitem [{\citenamefont {Akarsu}\ \emph {et~al.}(2019)\citenamefont {Akarsu}, \citenamefont {Kumar}, \citenamefont {Sharma},\ and\ \citenamefont {Tedesco}}]{Akarsu:2019pwn}%
  \BibitemOpen
  \bibfield  {author} {\bibinfo {author} {\bibfnamefont {{\"O}.}~\bibnamefont {Akarsu}}, \bibinfo {author} {\bibfnamefont {S.}~\bibnamefont {Kumar}}, \bibinfo {author} {\bibfnamefont {S.}~\bibnamefont {Sharma}},\ and\ \bibinfo {author} {\bibfnamefont {L.}~\bibnamefont {Tedesco}},\ }\bibfield  {title} {\bibinfo {title} {{Constraints on a Bianchi type I spacetime extension of the standard $\Lambda$CDM model}},\ }\href {https://doi.org/10.1103/PhysRevD.100.023532} {\bibfield  {journal} {\bibinfo  {journal} {Phys. Rev. D}\ }\textbf {\bibinfo {volume} {100}},\ \bibinfo {pages} {023532} (\bibinfo {year} {2019})},\ \Eprint {https://arxiv.org/abs/1905.06949} {arXiv:1905.06949 [astro-ph.CO]} \BibitemShut {NoStop}%
\bibitem [{\citenamefont {Aluri}\ \emph {et~al.}(2023)\citenamefont {Aluri} \emph {et~al.}}]{Aluri:2022hzs}%
  \BibitemOpen
  \bibfield  {author} {\bibinfo {author} {\bibfnamefont {P.~K.}\ \bibnamefont {Aluri}} \emph {et~al.},\ }\bibfield  {title} {\bibinfo {title} {{Is the observable Universe consistent with the cosmological principle?}},\ }\href {https://doi.org/10.1088/1361-6382/acbefc} {\bibfield  {journal} {\bibinfo  {journal} {Class. Quant. Grav.}\ }\textbf {\bibinfo {volume} {40}},\ \bibinfo {pages} {094001} (\bibinfo {year} {2023})},\ \Eprint {https://arxiv.org/abs/2207.05765} {arXiv:2207.05765 [astro-ph.CO]} \BibitemShut {NoStop}%
\bibitem [{\citenamefont {Hertzberg}\ and\ \citenamefont {Loeb}(2024)}]{Hertzberg:2024uqy}%
  \BibitemOpen
  \bibfield  {author} {\bibinfo {author} {\bibfnamefont {M.~P.}\ \bibnamefont {Hertzberg}}\ and\ \bibinfo {author} {\bibfnamefont {A.}~\bibnamefont {Loeb}},\ }\bibfield  {title} {\bibinfo {title} {{Constraints on an anisotropic universe}},\ }\href {https://doi.org/10.1103/PhysRevD.109.083538} {\bibfield  {journal} {\bibinfo  {journal} {Phys. Rev. D}\ }\textbf {\bibinfo {volume} {109}},\ \bibinfo {pages} {083538} (\bibinfo {year} {2024})},\ \Eprint {https://arxiv.org/abs/2401.15782} {arXiv:2401.15782 [astro-ph.CO]} \BibitemShut {NoStop}%
\bibitem [{\citenamefont {Akarsu}\ \emph {et~al.}(2023)\citenamefont {Akarsu}, \citenamefont {Di~Valentino}, \citenamefont {Kumar}, \citenamefont {Ozyigit},\ and\ \citenamefont {Sharma}}]{Akarsu:2021max}%
  \BibitemOpen
  \bibfield  {author} {\bibinfo {author} {\bibfnamefont {O.}~\bibnamefont {Akarsu}}, \bibinfo {author} {\bibfnamefont {E.}~\bibnamefont {Di~Valentino}}, \bibinfo {author} {\bibfnamefont {S.}~\bibnamefont {Kumar}}, \bibinfo {author} {\bibfnamefont {M.}~\bibnamefont {Ozyigit}},\ and\ \bibinfo {author} {\bibfnamefont {S.}~\bibnamefont {Sharma}},\ }\bibfield  {title} {\bibinfo {title} {{Testing spatial curvature and anisotropic expansion on top of the {\ensuremath{\Lambda}}CDM model}},\ }\href {https://doi.org/10.1016/j.dark.2022.101162} {\bibfield  {journal} {\bibinfo  {journal} {Phys. Dark Univ.}\ }\textbf {\bibinfo {volume} {39}},\ \bibinfo {pages} {101162} (\bibinfo {year} {2023})},\ \Eprint {https://arxiv.org/abs/2112.07807} {arXiv:2112.07807 [astro-ph.CO]} \BibitemShut {NoStop}%
\bibitem [{\citenamefont {Misner}(1969{\natexlab{a}})}]{Misner:1969hg}%
  \BibitemOpen
  \bibfield  {author} {\bibinfo {author} {\bibfnamefont {C.~W.}\ \bibnamefont {Misner}},\ }\bibfield  {title} {\bibinfo {title} {{Mixmaster universe}},\ }\href {https://doi.org/10.1103/PhysRevLett.22.1071} {\bibfield  {journal} {\bibinfo  {journal} {Phys. Rev. Lett.}\ }\textbf {\bibinfo {volume} {22}},\ \bibinfo {pages} {1071} (\bibinfo {year} {1969}{\natexlab{a}})}\BibitemShut {NoStop}%
\bibitem [{\citenamefont {Misner}(1969{\natexlab{b}})}]{Misner:1969ae}%
  \BibitemOpen
  \bibfield  {author} {\bibinfo {author} {\bibfnamefont {C.~W.}\ \bibnamefont {Misner}},\ }\bibfield  {title} {\bibinfo {title} {{Quantum cosmology. 1.}},\ }\href {https://doi.org/10.1103/PhysRev.186.1319} {\bibfield  {journal} {\bibinfo  {journal} {Phys. Rev.}\ }\textbf {\bibinfo {volume} {186}},\ \bibinfo {pages} {1319} (\bibinfo {year} {1969}{\natexlab{b}})}\BibitemShut {NoStop}%
\bibitem [{\citenamefont {{Belinski{\u{i}}}}\ \emph {et~al.}(1971)\citenamefont {{Belinski{\u{i}}}}, \citenamefont {{Lifshitz}},\ and\ \citenamefont {{Khalatnikov}}}]{1971SvPhU..13..745B}%
  \BibitemOpen
  \bibfield  {author} {\bibinfo {author} {\bibfnamefont {V.~A.}\ \bibnamefont {{Belinski{\u{i}}}}}, \bibinfo {author} {\bibfnamefont {E.~M.}\ \bibnamefont {{Lifshitz}}},\ and\ \bibinfo {author} {\bibfnamefont {I.~M.}\ \bibnamefont {{Khalatnikov}}},\ }\bibfield  {title} {\bibinfo {title} {{Reviews of Topical Problems: Oscillatory Approach to the Singular Point in Relativistic Cosmology}},\ }\href {https://doi.org/10.1070/PU1971v013n06ABEH004279} {\bibfield  {journal} {\bibinfo  {journal} {Soviet Physics Uspekhi}\ }\textbf {\bibinfo {volume} {13}},\ \bibinfo {pages} {745} (\bibinfo {year} {1971})}\BibitemShut {NoStop}%
\bibitem [{\citenamefont {Deliyergiyev}\ \emph {et~al.}(2025)\citenamefont {Deliyergiyev}, \citenamefont {Le~Delliou},\ and\ \citenamefont {Del~Popolo}}]{Deliyergiyev:2025kun}%
  \BibitemOpen
  \bibfield  {author} {\bibinfo {author} {\bibfnamefont {M.}~\bibnamefont {Deliyergiyev}}, \bibinfo {author} {\bibfnamefont {M.}~\bibnamefont {Le~Delliou}},\ and\ \bibinfo {author} {\bibfnamefont {A.}~\bibnamefont {Del~Popolo}},\ }\bibfield  {title} {\bibinfo {title} {{Hubble tension in an anisotropic Universe}},\ }\href {https://doi.org/10.1093/mnras/staf1374} {\bibfield  {journal} {\bibinfo  {journal} {Mon. Not. Roy. Astron. Soc.}\ }\textbf {\bibinfo {volume} {542}},\ \bibinfo {pages} {3105} (\bibinfo {year} {2025})},\ \Eprint {https://arxiv.org/abs/2510.19069} {arXiv:2510.19069 [astro-ph.CO]} \BibitemShut {NoStop}%
\bibitem [{\citenamefont {Ashtekar}\ and\ \citenamefont {Wilson-Ewing}(2009)}]{Ashtekar:2009vc}%
  \BibitemOpen
  \bibfield  {author} {\bibinfo {author} {\bibfnamefont {A.}~\bibnamefont {Ashtekar}}\ and\ \bibinfo {author} {\bibfnamefont {E.}~\bibnamefont {Wilson-Ewing}},\ }\bibfield  {title} {\bibinfo {title} {{Loop quantum cosmology of Bianchi I models}},\ }\href {https://doi.org/10.1103/PhysRevD.79.083535} {\bibfield  {journal} {\bibinfo  {journal} {Phys. Rev. D}\ }\textbf {\bibinfo {volume} {79}},\ \bibinfo {pages} {083535} (\bibinfo {year} {2009})},\ \Eprint {https://arxiv.org/abs/0903.3397} {arXiv:0903.3397 [gr-qc]} \BibitemShut {NoStop}%
\bibitem [{\citenamefont {de~Cesare}\ and\ \citenamefont {Wilson-Ewing}(2019)}]{deCesare:2019suk}%
  \BibitemOpen
  \bibfield  {author} {\bibinfo {author} {\bibfnamefont {M.}~\bibnamefont {de~Cesare}}\ and\ \bibinfo {author} {\bibfnamefont {E.}~\bibnamefont {Wilson-Ewing}},\ }\bibfield  {title} {\bibinfo {title} {{A generalized Kasner transition for bouncing Bianchi I models in modified gravity theories}},\ }\href {https://doi.org/10.1088/1475-7516/2019/12/039} {\bibfield  {journal} {\bibinfo  {journal} {JCAP}\ }\textbf {\bibinfo {volume} {12}},\ \bibinfo {pages} {039}},\ \Eprint {https://arxiv.org/abs/1910.03616} {arXiv:1910.03616 [gr-qc]} \BibitemShut {NoStop}%
\bibitem [{\citenamefont {Motaharfar}\ \emph {et~al.}(2024)\citenamefont {Motaharfar}, \citenamefont {Singh},\ and\ \citenamefont {Thareja}}]{Motaharfar:2023hil}%
  \BibitemOpen
  \bibfield  {author} {\bibinfo {author} {\bibfnamefont {M.}~\bibnamefont {Motaharfar}}, \bibinfo {author} {\bibfnamefont {P.}~\bibnamefont {Singh}},\ and\ \bibinfo {author} {\bibfnamefont {E.}~\bibnamefont {Thareja}},\ }\bibfield  {title} {\bibinfo {title} {{Classicality and uniqueness in the loop quantization of Bianchi I spacetimes}},\ }\href {https://doi.org/10.1103/PhysRevD.109.086013} {\bibfield  {journal} {\bibinfo  {journal} {Phys. Rev. D}\ }\textbf {\bibinfo {volume} {109}},\ \bibinfo {pages} {086013} (\bibinfo {year} {2024})},\ \Eprint {https://arxiv.org/abs/2311.08465} {arXiv:2311.08465 [gr-qc]} \BibitemShut {NoStop}%
\bibitem [{\citenamefont {McNamara}\ \emph {et~al.}(2023)\citenamefont {McNamara}, \citenamefont {Saini},\ and\ \citenamefont {Singh}}]{McNamara:2022dmf}%
  \BibitemOpen
  \bibfield  {author} {\bibinfo {author} {\bibfnamefont {A.~M.}\ \bibnamefont {McNamara}}, \bibinfo {author} {\bibfnamefont {S.}~\bibnamefont {Saini}},\ and\ \bibinfo {author} {\bibfnamefont {P.}~\bibnamefont {Singh}},\ }\bibfield  {title} {\bibinfo {title} {{Novel relationship between shear and energy density at the bounce in nonsingular Bianchi I spacetimes}},\ }\href {https://doi.org/10.1103/PhysRevD.107.026003} {\bibfield  {journal} {\bibinfo  {journal} {Phys. Rev. D}\ }\textbf {\bibinfo {volume} {107}},\ \bibinfo {pages} {026003} (\bibinfo {year} {2023})},\ \Eprint {https://arxiv.org/abs/2210.07257} {arXiv:2210.07257 [gr-qc]} \BibitemShut {NoStop}%
\bibitem [{\citenamefont {Zulfiqar}\ and\ \citenamefont {Hassan}(2025)}]{Zulfiqar:2025chv}%
  \BibitemOpen
  \bibfield  {author} {\bibinfo {author} {\bibfnamefont {A.}~\bibnamefont {Zulfiqar}}\ and\ \bibinfo {author} {\bibfnamefont {S.~M.}\ \bibnamefont {Hassan}},\ }\bibfield  {title} {\bibinfo {title} {{Polymer Bianchi-I with polymer matter}},\ }\href@noop {} {\  (\bibinfo {year} {2025})},\ \Eprint {https://arxiv.org/abs/2509.24586} {arXiv:2509.24586 [gr-qc]} \BibitemShut {NoStop}%
\bibitem [{\citenamefont {Barca}\ \emph {et~al.}(2022)\citenamefont {Barca}, \citenamefont {Giovannetti},\ and\ \citenamefont {Montani}}]{Barca:2021epy}%
  \BibitemOpen
  \bibfield  {author} {\bibinfo {author} {\bibfnamefont {G.}~\bibnamefont {Barca}}, \bibinfo {author} {\bibfnamefont {E.}~\bibnamefont {Giovannetti}},\ and\ \bibinfo {author} {\bibfnamefont {G.}~\bibnamefont {Montani}},\ }\bibfield  {title} {\bibinfo {title} {{Comparison of the semiclassical and quantum dynamics of the Bianchi I cosmology in the polymer and GUP extended paradigms}},\ }\href {https://doi.org/10.1142/S0219887822500979} {\bibfield  {journal} {\bibinfo  {journal} {Int. J. Geom. Meth. Mod. Phys.}\ }\textbf {\bibinfo {volume} {19}},\ \bibinfo {pages} {2250097} (\bibinfo {year} {2022})},\ \Eprint {https://arxiv.org/abs/2112.08905} {arXiv:2112.08905 [gr-qc]} \BibitemShut {NoStop}%
\bibitem [{\citenamefont {Barca}\ and\ \citenamefont {Gielen}(2025)}]{Barca:2025fpu}%
  \BibitemOpen
  \bibfield  {author} {\bibinfo {author} {\bibfnamefont {G.}~\bibnamefont {Barca}}\ and\ \bibinfo {author} {\bibfnamefont {S.}~\bibnamefont {Gielen}},\ }\bibfield  {title} {\bibinfo {title} {{Bouncing Bianchi models with deformed commutation relations}},\ }\href {https://doi.org/10.1103/fwfy-gtgx} {\bibfield  {journal} {\bibinfo  {journal} {Phys. Rev. D}\ }\textbf {\bibinfo {volume} {112}},\ \bibinfo {pages} {086014} (\bibinfo {year} {2025})},\ \Eprint {https://arxiv.org/abs/2507.01678} {arXiv:2507.01678 [gr-qc]} \BibitemShut {NoStop}%
\bibitem [{\citenamefont {Linsefors}\ and\ \citenamefont {Barrau}(2014)}]{Linsefors:2013bua}%
  \BibitemOpen
  \bibfield  {author} {\bibinfo {author} {\bibfnamefont {L.}~\bibnamefont {Linsefors}}\ and\ \bibinfo {author} {\bibfnamefont {A.}~\bibnamefont {Barrau}},\ }\bibfield  {title} {\bibinfo {title} {{Modified Friedmann equation and survey of solutions in effective Bianchi-I loop quantum cosmology}},\ }\href {https://doi.org/10.1088/0264-9381/31/1/015018} {\bibfield  {journal} {\bibinfo  {journal} {Class. Quant. Grav.}\ }\textbf {\bibinfo {volume} {31}},\ \bibinfo {pages} {015018} (\bibinfo {year} {2014})},\ \Eprint {https://arxiv.org/abs/1305.4516} {arXiv:1305.4516 [gr-qc]} \BibitemShut {NoStop}%
\bibitem [{\citenamefont {Chiou}\ and\ \citenamefont {Vandersloot}(2007)}]{Chiou:2007sp}%
  \BibitemOpen
  \bibfield  {author} {\bibinfo {author} {\bibfnamefont {D.-W.}\ \bibnamefont {Chiou}}\ and\ \bibinfo {author} {\bibfnamefont {K.}~\bibnamefont {Vandersloot}},\ }\bibfield  {title} {\bibinfo {title} {{The Behavior of non-linear anisotropies in bouncing Bianchi I models of loop quantum cosmology}},\ }\href {https://doi.org/10.1103/PhysRevD.76.084015} {\bibfield  {journal} {\bibinfo  {journal} {Phys. Rev. D}\ }\textbf {\bibinfo {volume} {76}},\ \bibinfo {pages} {084015} (\bibinfo {year} {2007})},\ \Eprint {https://arxiv.org/abs/0707.2548} {arXiv:0707.2548 [gr-qc]} \BibitemShut {NoStop}%
\end{thebibliography}%

\end{document}